\documentclass[twocolumn,showpacs,prd]{revtex4}
\usepackage{epsfig}

\begin{document}

\title{Probing the 2-3 leptonic mixing at high-energy
neutrino telescopes}
\author{Pasquale D.~Serpico}
\affiliation{Max-Planck-Institut f\"ur Physik
(Werner-Heisenberg-Institut), F\"ohringer Ring 6, 80805 M\"unchen,
Germany}

\date{\today}

\begin{abstract}
We discuss the possibility to probe leptonic mixing parameters at high-energy
neutrino telescopes in a model-independent way, using astrophysical
neutron and pion sources. In particular we show how the octant of
the 2-3 mixing angle might be determined independently of prior knowledge 
of the source, even when current uncertainties on the other mixing parameters 
are included. We also argue that non-trivial neutrino oscillation effects 
should be taken into account when using high-energy flavor ratios for 
astrophysical diagnostics.
\end{abstract}

\pacs{14.60.Lm, 
14.60.Pq,       
95.85.Ry        
\hfill Preprint MPP-2005-154}

\maketitle

\section{Introduction}\label{introduction}
The physics of neutrinos is entering the precision era, where
laboratory sources from reactors or accelerators are progressively
replacing the natural ones (solar and atmospheric neutrinos) for the
detailed determination of mixing angles and mass splittings. A
parallel development is expected with the opening of a new
observational window in astrophysics by the next generation of
high-energy neutrino telescopes~\cite{Learned:2000sw}. The hope is
that these instruments will finally shed light on several open
problems in cosmic-ray and gamma-ray astrophysics. It is
well recognized that the value of neutrino oscillation
parameters has a great impact on the expected fluxes. We just
mention here that significant $\nu_\tau$ fluxes from astrophysical
sources are implied by the large mixing in the
$\nu_\mu$--$\nu_\tau$ sector, opening the interesting possibility to
identify $\nu_\tau$ events in both optical Cherenkov telescopes and
in extensive air-shower 
experiments~\cite{Fargion:1997eg,Dutta:2000jv,Athar:2002rr}.
On the other hand, high-energy neutrinos are expected to come mainly
from pion decays, with a flavor composition of
$\{\phi_e:\phi_\mu:\phi_\tau\}\simeq\{\frac{1}{3}:\frac{2}{3}:0\}$ at
the source. (We denote with $\phi_\alpha$ the combined flux
of $\nu_\alpha$ and $\bar\nu_\alpha$, where $\alpha=e,\mu,\tau$.)
Presently favored ranges for mixing parameters imply oscillated
fluxes at the detector $\phi_{\alpha}^D$ approximately in the ratios
$\{\frac{1}{3}:\frac{1}{3}:\frac{1}{3}\}$, almost independently from
the details of neutrino mixing parameters~\cite{Athar:2000yw,Barenboim:2003jm}. Astrophysical
uncertainties and the expected low statistics would not justify
deeper studies. This qualitative argument probably explains why
 the potential for neutrino mixing studies has remained
largely unexplored.

However, the possibility to exploit the flavor content of neutrino
fluxes for astrophysical diagnostics has been recently analyzed in
greater detail. A change in neutrino flavor fluxes was 
shown to be important for diagnostics of gamma ray bursts
(GRB)~\cite{Kashti:2005qa}. As another example, the fraction of
neutrinos produced in high-energy accelerators via the strongly
isospin-asymmetric $p\gamma$ process (as opposed to $pp$ inelastic
scattering) might also be measurable, at least around energies of
6.3 PeV (~1~PeV$\equiv\,$10$^{15}\,$eV)~\cite{Anchordoqui:2004eb,BG}. This idea
is based on the fact that at this energy optical Cherenkov
telescopes have an enhanced efficiency to single out
$\bar{\nu}_e$ showers, because of the Glashow resonant process
$\bar\nu_e+e^{-}\to W^{-}$. The ratio $R_G$ of such events to the
$\nu_\mu$ plus $\bar\nu_\mu$ charged current tracks in the same
energy bin is a suitable observable to that purpose. The ratio $R_G$
was shown to have a significant sensitivity to the mixing angles as
well, in particular to the 1-2 mixing angle $\theta_{12}$~\cite{BG}. 
In that respect, we
also have discussed an interesting astrophysical
target~\cite{Serpico:2005sz}. Neutron primaries generated in the
photo-dissociation of nuclei (see e.g.~\cite{Anchordoqui:2003vc})
would have the right properties to explain the excess of high-energy
cosmic rays from the Galactic Plane at $E\simeq 10^{18}$~eV, reported
in~\cite{anisotropy}.
If this model is correct, the initial flavor content of the neutrino
flux from some galactic regions is close to $\{1:0:0\}$. 
A sensitivity to $\theta_{13}$ and to the
leptonic CP-phase $\delta_{\rm CP}$ is achieved via 
the quantity $R\equiv\phi_\mu^D/(\phi_e^D+\phi_\tau^D)$,
that can be deduced in a neutrino telescope from the ratio of 
track to shower
events~\cite{Beacom:2003nh}~\footnote{The quantity $R$ approaches the ratio 
of track to shower events in the limit in which sub-leading 
neutral-currrent events are neglected, and if both channels 
are detected with the same efficiency. Typical size of track 
to shower ratio is of several,
since the km-long muon range ensures a larger effective 
target volume for the tracks.
This is however a technical point inessential to our
considerations, since it could be accounted for in refined
predictions and for a specific experimental setup.}. 
However, though physically plausible, this source is not
guaranteed. The very existence of a significant anisotropy---at
least towards the Galactic Center---is currently debated, in the
light of the negative results of an analysis of preliminary Auger
data~\cite{Letessier-Selvon:2005uf}.

In the following, we generalize previous considerations arguing
that: (i) There could be neutron beam sources invisible to cosmic
ray (an-)isotropy observations, and only detectable indirectly at
neutrino telescopes. (ii)~Other candidate targets useful for
neutrino mixing studies at neutrino telescopes also exist, like
muon-damped $\nu_\mu$ sources from pion decays, that were recently 
discussed~\cite{Kashti:2005qa}. We shall motivate that both classes of
sources could be not only identified at neutrino telescopes, but
also used to infer non-trivial information on certain neutrino mixing
parameters in a {\it model-independent way}, i.e. irrespective of
astrophysical uncertainties. The argument still holds when presently
allowed ranges for the other mixing parameters are taken into account. In
particular, we shall show how a robust lower bound on $\theta_{23}$ 
could be established, and thus a value of
$\theta_{23}>45^\circ$ identified. Note that this information 
is non-trivial. The present 2$\,\sigma$ 
range is $36^\circ\leq\theta_{23}\leq
52^\circ$~\cite{Fogli:2005cq} and a deviation from maximal 
mixing would be important for flavor symmetries
and neutrino mass models~\cite{Dorsner:2004qb}.

Of course, our considerations could be invalidated if
exotic mechanisms like neutrino decays are
effective~\cite{Beacom:2002vi}, but such scenarios seem to be at 
least disfavored by cosmological bounds~\cite{Hannestad:2005ex,Bell:2005dr}.

In Section~\ref{extension} we treat generic neutron beam sources and
focus on the octant of $\theta_{23}$ as a model-independent
parameter possibly accessible at neutrino telescopes. In
Section~\ref{mudamp} similar considerations are developed for a pure
$\nu_{\mu}$ beam from pion decay. In Section \ref{concl} we
conclude.

\section{Neutron beam sources}\label{extension}
Neutrino fluxes detectable at neutrino telescopes, i.e. at
$E\agt\,$0.1--1~TeV, might well originate in the decay of few~PeV
neutrons from sources which have characteristics similar to the ones
detailed in~\cite{Anchordoqui:2003vc}, but whose neutron spectrum
cuts-off at energies $E\ll 10^{18}$ eV. Since the decay length of a
neutron is $d_n\simeq\,$10 pc $(E_n/{\rm PeV})$, and typical galactic
distances are of order 10 kpc, such a source would not show up as a
cosmic ray anisotropy. 
\begin{figure}[!thb]
\centering
\epsfig{file=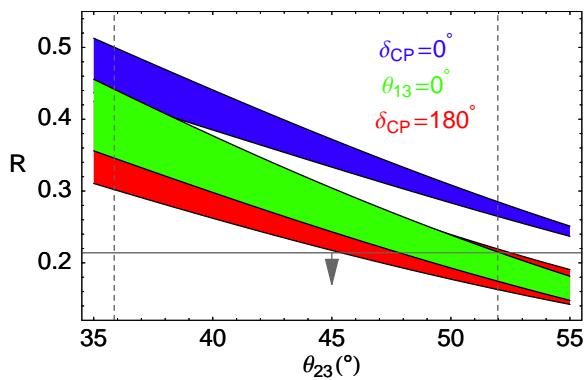,width=\columnwidth}
\caption{The ratio $R$ of Eq.~(\ref{ratioNbeam}) for a neutron beam 
source vs.~$\theta_{23}$.
The bands show the effect of the uncertainty on $\theta_{12}$, and
the trivial case $\{\theta_{13}=0^\circ\}$ is plotted together with
the limiting cases $\{\theta_{13}=10.3^\circ,\delta_{\rm
CP}=0^\circ\}$ and $\{\theta_{13}=10.3^\circ,\delta_{\rm
CP}=180^\circ\}$. Dashed vertical lines enclose the allowed range for
$\theta_{23}$. The ranges shown are the 95\% C.L. according 
to~\cite{Fogli:2005cq}. The region below the solid horizontal line
requires $\theta_{23}>45^\circ$.}\label{octantNB}
\end{figure}
The decay protons would rapidly lose
directional information via deflection in the galactic magnetic
field. The dominance of nuclei in the galactic cosmic ray spectrum
is likely starting just above $10^{15}\,$eV, and the spectrum of
galactic cosmic rays is expected to extend at least up to
few$\,\times 10^{17}\,$eV (see e.g.~\cite{Berezinsky:2004wx} and
Section 3.1 in~\cite{Kachelriess:2005qm}): A situation suitable to
the neutron beam production is conceivable in many regions of our
Galaxy. Of course, such ``hidden'' neutron beams could only be
revealed by neutrino observations, and are thus constrained only by
the direct observational upper bounds at neutrino telescopes.
Nonetheless, any standard pion-decay source would produce
neutrinos with a ratio $R\simeq 0.5$ (or larger, see
Section~\ref{mudamp}). By observing a ratio $R$ significantly lower
than 0.5, one could claim both the discovery of an invisible neutron
beam and put constraints on the neutrino mixing parameters,
since any ``background'' (i.e., any additional flux not sharing the same
flavor content) should push $R$ to higher values.
More quantitatively, the flux flavor ratios $\phi_\beta^D$
arriving at the detector are given in terms of the probabilities
$P_{\alpha\beta}\equiv P(\nu_\alpha\to\nu_\beta)$ as $\phi^D_\beta
=\sum_\alpha P_{\alpha\beta} \phi_\alpha$, where $\phi_\alpha$ are
the flux ratios at the source. Matter effects are negligible because
of the extremely low densities of cosmic environments, and the
interference terms sensitive to the mass splittings and to the sign
of $\delta_{\rm CP}$ (i.e., the CP-violating terms) average out
because the galactic distances far exceed the experimentally
known oscillation lengths. This also
implies that the same probability formulae apply to neutrino and
anti-neutrino channels. One then obtains
\begin{equation}
  P_{\alpha\beta}=\delta_{\alpha \beta}-2\sum_{j>k}{\rm Re}(U_{\beta j}
  U_{\beta k}^{*} U_{\alpha j}^{*} U_{\alpha k}) \,,\label{exacteq}
\end{equation}
where $U(\theta_{12},\theta_{23},\theta_{13},\delta_{\rm CP})$ is the neutrino mixing matrix~\cite{Eidelman:2004wy} and greek (latin) letters are used as 
flavor (mass) indices.
For a neutron beam source, the parameter $R$ can be
expressed in terms of oscillation probabilities $P_{\alpha\beta}$ as
\begin{equation}
R\equiv\frac{\phi_\mu^D}{\phi_e^D+\phi_\tau^D}=\frac{P_{e\mu}}{P_{ee}+P_{e\tau}}=\frac{P_{e\mu}}{1-P_{e\mu}}\label{ratioNbeam},
\end{equation}
the last equality following from unitarity, $P_{ee}+P_{e\mu}+P_{e\tau}=1$.
An useful approximation for $P_{e\mu}$, obtained as first
order expansion in the small quantity $\sin\theta_{13}$, is
\begin{widetext}
\begin{equation}
P_{e\mu}
=\frac{1}{2}\sin^22\theta_{12}\cos^2\theta_{23}+\frac{1}{2}\sin\theta_{13}\cos\delta_{\rm CP}\sin2\theta_{12}\cos2\theta_{12}\sin2\theta_{23}+{\cal O}(\theta_{13}^2).\label{nbeam}
\end{equation}
\end{widetext}
The extremes of the ratio Eq.~(\ref{ratioNbeam}) for a fixed $\theta_{12}$ and $\theta_{23}$
are obtained for maximal allowed value of $\theta_{13}$ and the cases
$\cos\delta_{\rm CP}=\pm 1$, as the linear approximation of
Eq.~(\ref{nbeam}) suggests. In Fig.~\ref{octantNB} it is clearly
shown that, also including current 2$\,\sigma$ uncertainties on
mixing parameters, observations of an extremely low value for $R$,
say $R\alt 0.21$, could only be reconciled with $\theta_{23}>45^\circ$.

Until now we have focused on neutron beams from galactic sources,
because they are motivated targets having a chance of detection. On the
other hand, it should be stressed that extra-galactic sources that
have suitable conditions also exist. Even when one turns to the most
reliable of the extra-galactic neutrino sources, the cosmogenic
neutrino flux, one easily realizes that at energies around
$10^{16}\,$eV a secondary peak almost purely made of $\bar\nu_e$
should be present~\cite{Engel:2001hd}. This is formed after neutron
decays, both in the case of proton and heavy nuclei primaries. In
the latter case a relatively larger contribution is expected because
of the additional free-neutrons produced in
photo-dissociations~\cite{Hooper:2004jc,Ave:2004uj}. Of course, this
flux is so low that a detection is challenging, and maybe prevented
even in principle by the larger contributions from ``canonical''
diffuse fluxes from other extra-galactic sources.

\section{Muon damped sources}\label{mudamp}
We now turn to another class of sources producing a
non-trivial flavor content at neutrino telescopes, i.e.
sources optically thick to muons (lifetime $\simeq 2.2\times
10^{-6}\,$s) but not to pions (lifetime $\simeq 2.6\times
10^{-8}\,$s), which would mainly emit neutrinos in the flavor ratios
$\{0:1:0\}$. Such flavor content arises in specific astrophysical
models~\cite{Rachen:1998fd}. In addition, for any concrete example
of accelerating engine, a transition from
$\{\frac{1}{3}:\frac{2}{3}:0\}$ to $\{0:1:0\}$ is expected in some
energy range. This follows from the competition between growing
decay path-length ($\lambda_{\rm decay}\propto E$) and decreasing
energy-loss distance with energy. For standard active galactic
nuclei models the transition energy is expected to be quite high,
around $10^6$ TeV, but it might be as low as $\sim$10 TeV for
GRB. Such a phenomenon could also offer interesting perspectives for
GRB diagnostics~\cite{Kashti:2005qa}. Again, the effect of mixing
angles in this case is non-trivial, since the flavor ratio at the
Earth for a $\{0:1:0\}$ source would depend only on the $\nu_\mu$
survival probability as
\begin{equation}
R\equiv\frac{\phi_\mu^D}{\phi_e^D+\phi_\tau^D}=\frac{P_{\mu\mu}}{1-P_{\mu\mu}},
\end{equation}
where we used the unitarity condition
 $P_{\mu
e}+P_{\mu\tau}=1-P_{\mu\mu}$. For illustrative purpose we report
here a first order expansion of $P_{\mu\mu}$ in the small quantity
$\sin\theta_{13}$,
\begin{widetext}
\begin{equation}
P_{\mu\mu}= 1-2\cos^2\theta_{23}
\left[\sin^2\theta_{23}+\frac{1}{4}\sin^22\theta_{12}\cos^2\theta_{23}
+\frac{1}{2}\sin\theta_{13}\cos\delta_{\rm CP}\sin2\theta_{12}\cos2\theta_{12}\sin2\theta_{23}\right]+{\cal O}(\theta_{13}^2).\label{pbeam}
\end{equation}
\end{widetext}

\begin{figure}
\centering
\epsfig{file=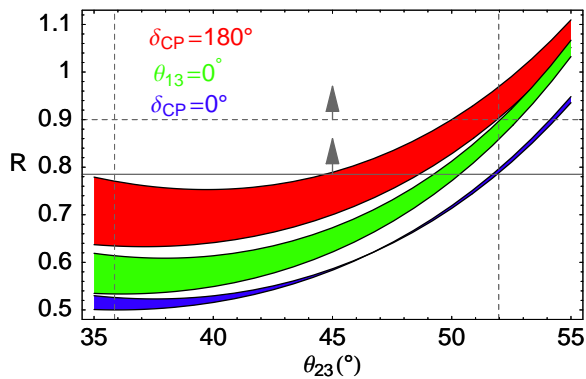,width=\columnwidth}
\caption{Same as Fig.~\ref{octantNB}, for a source optically thick to
muons, but not to pions. The region above the solid horizontal line
requires $\theta_{23}>45^\circ$, the one above the dashed horizontal
line in addition requires a non-vanishing $\{\theta_{13},\,\delta_{\rm CP}\}$
sector.} \label{thicknu}
\end{figure}

In Fig.~\ref{thicknu} we show that any observation of a ratio
$R\agt 0.78$ would not only point to a muon-damped source,
but would also constrain the octant of $\theta_{23}$, i.e.
$\theta_{23}>45^\circ$. This result is irrespective of the
uncertainties on the other mixing parameters, as well as of known
backgrounds from (undamped) pion chain or even neutron beams, which
could only contribute to lower the value of $R$. Note also
that in both Figs.~\ref{octantNB} and~\ref{thicknu}, special
regions in the parameter space exist that are only compatible with
very specific values of the mixing parameters, and in particular
$\theta_{13}$ and $\delta_{\rm CP}$. For example, if one establishes
independently that $\theta_{23}<45^\circ$, a detected value of
$R\agt 0.7$ would only be compatible with a relatively large
$\theta_{13}$ and a non-vanishing $\delta_{\rm CP}$.

The prospects of forthcoming laboratory experiments for 
the determination of the octant of $\theta_{23}$ 
have been analyzed in~\cite{Antusch:2004yx},
and more in general for neutrino oscillation parameters 
in~\cite{Huber:2004ug}. To give a quantitative example of the possibilities 
of neutrino telescopes, for a benchmark flux of muon 
neutrinos of $E^2 {\rm d}N/{\rm d}E= 10^{-7}\,$GeV cm$^{-2}$ s$^{-1}$, 
IceCube~\cite{Ahrens:2002dv} could determine 
the flavor ratio at the 15\% level after 1 year~\cite{Beacom:2003nh}.
In this case, IceCube would be sufficiently sensitive to detect the 
effects described above. 
Moderately lower fluxes could be compensated by a 
larger integration time. Remarkably, in any kind of extraterrestrial 
flux, even a diffuse one, one could look for an energy range 
with peculiar flavor ratios. Moreover, adding complementary
information from forthcoming laboratory experiments would improve
the chances to identify e.g. the effect of $\theta_{23}>45^\circ$  at
neutrino telescopes.

\section{Conclusion}\label{concl}
We have argued that the role of neutrino mixing at high-energy
neutrino telescopes is not trivial, and that forthcoming
observations are potentially interesting for neutrino mixing
phenomenology. If sources like neutron or pion beams exist, we
showed that: (i)~They can be identified unambiguously at neutrino
telescopes; (ii)~They  may allow a model-independent determination of
crucial qualitative features of neutrino mixing parameters, like the
octant of $\theta_{23}$ or the existence of a non-vanishing
$\{\theta_{13},\,\delta_{\rm CP}\}$ sector.

From a complementary perspective, accurate laboratory measurements 
of neutrino mixing parameters are of primary
importance to perform astrophysical diagnostics: Since the flux
flavor ratios depend on mixing angles, degeneracies with
astrophysical parameters may arise. For example, although the main
emphasis in~\cite{BG} was on the sensitivity of the ratio $R_G$ to
$\theta_{12}$ (which is relatively well determined from solar
neutrino experiments), we remark that varying $\theta_{13}$ in the
allowed experimental range can have an impact as large as $\simeq
15\%$ on $R_G$. This effect alone might affect
the extraction of astrophysical parameters.

Throughout this paper, we have conservatively assumed that only the
ratio $R$ can be determined at neutrino telescopes. At energies larger
than a few PeV, and in particular around 6.3 PeV where the
observable $R_G$ can be used, one might expect to measure or to
constrain the $\tau$ flavor fraction as well, since
$\nu_\tau$-specific signatures such as lolly-pop or double bang
events can be detected~\cite{Beacom:2003nh}. It is clear that the
chance for a multi-channel observation offers a more powerful tool.

We conclude that the usual assumption of a canonical
flavor equipartition at neutrino telescopes is too simplistic:
Peculiar astrophysical sources may offer complementary constraints
to laboratory measurements or, conversely, a more accurate
experimental determination of mixing parameters may help to shed 
light on the properties of cosmic accelerators. After the pioneering 
era of the discovery of the solar neutrino deficit and of the atmospheric
neutrino anomaly, observations at the highest energies will be sensitive 
to new astrophysical sources, that might still offer opportunities for 
neutrino oscillation studies.
\section*{Acknowledgements} 
The author thanks M. Kachelrie{\ss}, A. Mirizzi, and G. Raffelt for
reading the manuscript and for useful comments, and acknowledges 
the support by the Deut\-sche For\-schungs\-ge\-mein\-schaft under grant
SFB 375 and by the European Network of Theoretical Astroparticle Physics 
ILIAS/N6 under contract number RII3-CT-2004-506222.

\end{document}